\documentclass{mem}
\usepackage{graphicx}
\usepackage{txfonts,epsfig,graphicx,natbib,url,twoopt}
\usepackage[breaklinks=true]{hyperref} 
\usepackage[labelfont=footnotesize,textfont=footnotesize]{caption}   

\bibpunct{(}{)}{;}{a}{}{,}    
\newcommandtwoopt{\citeads}[3][][]{\href{http://adsabs.harvard.edu/abs/#3}%
                                        {\citealp[#1][#2]{#3}}}
\newcommandtwoopt{\citepads}[3][][]{\href{http://adsabs.harvard.edu/abs/#3}%
                                         {\citep[#1][#2]{#3}}}
\newcommandtwoopt{\citetads}[3][][]{\href{http://adsabs.harvard.edu/abs/#3}%
                                         {\citet[#1][#2]{#3}}}
\newcommandtwoopt{\citeyearads}
  [3][][]{\href{http://adsabs.harvard.edu/abs/#3}%
  {\citeyear[#1][#2]{#3}}}
\begin{document}
\def\teff{$T\rm_{eff }$}
\def\kms{$\mathrm {km s}^{-1}$}

\title{
Balloon-borne hard X-ray astronomy with PoGOLite: Opening a new window on the universe
}

   \subtitle{}

\author{
M. Jackson\inst{1}, on behalf of the PoGOLite collaboration
          }

  \offprints{M. Jackson}

\institute{
KTH Royal Institute of Technology, Department of Physics and The Oskar Klein Centre for Cosmoparticle Physics,
AlbaNova University Centre,
10691 
Stockholm, Sweden. 
\email{miranda@particle.kth.se}
}

\authorrunning{Jackson }

\titlerunning{PoGOLite}

\abstract{
High energy polarization can be an indication of geometry, orientation, and other physical 
phenomena for a variety of sources, but has heretofore been a 
virtually unmeasured phenomenon. PoGOLite is a balloon-borne instrument intended to 
measure the polarization of X-rays between 25 and 100 keV. The pathfinder version of the 
instrument is scheduled
to be launched from northern Sweden in summer 2013. The primary targets are 
the Crab pulsar and nebula and Cygnus X-1.

\keywords{
Instrumentation: polarimeters -- Crab: pulsars: individual -- 
ISM: supernova remnants -- X-rays: binaries -- Polarization}
}
\maketitle{}

\section{Introduction}
PoGOLite \citepads{2008APh....30...72K} is a Compton-based X-ray polarimeter capable of measuring the polarization of X-rays with energies between 25 and 100 keV. This paper describes the pathfinder instrument which has a smaller effective area than the full instrument and is scheduled to be launched in 2013. 

\section{Detector array and electronics}
The detector array comprises 61 hexagonal phoswich detector cells (PDC) and 30 side anticoincidence shield (SAS) units composed of BGO material. A polarization event occurs when a photon Compton scatters in one fast plastic scintillator and is absorbed in another fast scintillator.

The polarimeter electronics consist of several multichannel flash-ADC boards and other boards which combine the signals and perform logical functions. Because of the active and passive collimation employed in the detector array, as well as the rudimentary pulse shape and height analysis employed in the electronics, the instrument is optimized for point sources, and most background events will be rejected immediately, resulting in lower dead time. It is possible to set parameters such as photomultiplier tube voltage and various thresholds for each channel, in order to make the responses as uniform as possible.

A neutron detector \citep{tak10} will provide a measurement of the neutron background in flight. Neutrons are the largest source of background for PoGOLite, and can even simulate polarization events, so it is important to understand the neutron background as thoroughly as possible.

\section{Pointing system, software and operations}
PoGOLite employs a sophisticated gimbaled pointing system with differential GPS and star trackers used for the fine pointing. The pointing accuracy is better than 0.1 degree, which ensures that the signal will be optimal \citep{cmb10}. The polarimeter is rotated about its axis once for each observation, which largely negates any systematic effects caused by varying detector responses.

After a brief period at the beginning of the flight where a high speed connection is possible, a slower Iridium (www.iridium.com, accessed 2012 Nov 21) connection must be used. It will be impossible to transfer the data to a ground computer during the Iridium phase of the flight, so the data must be stored onboard for later retrieval. Redundant storage is provided by three PC104 computers with RAID SSD arrays. One of these computers is contained within a specially made ``black box" enclosure which is designed to damp the shock of parachute opening and the impact with the ground at landing. In addition, a significant amount of preprocessing is done on the data, and the resulting event lists are saved in a compact form which can be transferred to the ground during Iridium operation. 

The software in PoGOLite is designed to automatically observe the optimal target based on the time, date, and location. This ensures that the instrument will function and collect data even when no connection from the ground is possible. The primary software provides an automatic synchronization between pointing system operations and polarimeter functions, which ensures that the instrument is accurately pointing throughout each observation.

\section{Flight parameters and targets}

The payload has a mass of approximately two tonnes, and the balloon and flight train add an additional two tonnes. The balloon is capable of lifting such a mass to an altitude of above 38.5 km. Daily fluctuations in temperature will result in increases and decreases in the altitude. The signal increases significantly at higher altitudes, so it is desired that the highest possible altitude be maintained for as long as possible during the flight.

The instrument, in its present form, was launched in 2011 July from Esrange in northern Sweden. However, due to a balloon malfunction, the flight was terminated before the final altitude was reached. In 2012, the presence of surface winds prevented a launch attempt. An approximately 5-day flight is expected to take place in summer 2013 from Esrange to northern Canada. It is also possible that the flight will be circumpolar, which would increase the duration of the flight to 17 days or more, in which case the payload would return to the ground near the launch site.

For the maiden flight, the Crab pulsar and nebula will be observed as much as possible in order to maximize the statistics. When the Crab is too low on the sky, Cygnus X-1 will be the primary target. It may be possible to observe solar flares or flaring objects during the flight, as the instrument can be pointed at will.

\section{Conclusions}

PoGOLite holds great promise to provide a first look at the X-ray polarization of the Crab nebula and Cygnus X-1 with a polarimeter. It may be possible to produce a polarization lightcurve of the Crab pulsar of a quality which will provide a test of pulsar models, such as the caustic and outer gap models. Measurements of X-ray polarization such as those made with PoGOLite will provide a new way of looking at extreme objects.

\bibliographystyle{aa}
\bibliography{jacksonm}

\begin{thebibliography}{}

\bibitem[{{Kamae} {et al.}(2008){Kamae}, T. and {Andersson}, V. and {Arimoto}, M. and {Axelsson}, M. and 
	{Marini Bettolo}, C. and {Bj{\"o}rnsson}, C.-I. and {Bogaert}, G. and 
	{Carlson}, P. and {Craig}, W. and {Ekeberg}, T. and {Engdeg{\aa}rd}, O. and 
	{Fukazawa}, Y. and {Gunji}, S. and {Hjalmarsdotter}, L. and 
	{Iwan}, B. and {Kanai}, Y. and {Kataoka}, J. and {Kawai}, N. and 
	{Kazejev}, J. and {Kiss}, M. and {Klamra}, W. and {Larsson}, S. and 
	{Madejski}, G. and {Mizuno}, T. and {Ng}, J. and {Pearce}, M. and 
	{Ryde}, F. and {Suhonen}, M. and {Tajima}, H. and {Takahashi}, H. and 
	{Takahashi}, T. and {Tanaka}, T. and {Thurston}, T. and {Ueno}, M. and 
	{Varner}, G. and {Yamamoto}, K. and {Yamashita}, Y. and {Ylinen}, T. and 
	{Yoshida}, H.}]{2008APh....30...72K} Kamae, T., et al. {\it Astroparticle Physics}  30 (2008) 72.
 
 
\bibitem[Takahashi et al. (2010)]{tak10} Takahashi, H., et al. 
{\it Nuclear Science Symposium Conference Record (NSS/MIC)}, 2010 IEEE, pp32-37.



\bibitem[Marini Bettolo (2010)]{cmb10}  Marini Bettolo, C. (2010).
                              \textit{Performance studies and star tracking for PoGOLite}, KTH Doctoral thesis, Stockholm, Sweden, 139.

\end{thebibliography}

\end{document}